\documentclass[journal=jpclcd,manuscript=letter]{achemso}

\usepackage{amsmath}
\usepackage{amssymb}
\usepackage{bm}
\usepackage{multirow}
\usepackage{subcaption}
\usepackage{tikz}
\usepackage{comment}

\renewcommand{\phi}{\ensuremath{\varphi}}

\newcommand{\sumover}[2]{\ensuremath{\underset{#1}{\overset{#2}{\sum}}}}
\newcommand{\bcreation}[1]{\ensuremath{b_{#1}^\dagger}}
\newcommand{\bannihilation}[1]{\ensuremath{b_{#1}}}
\newcommand{\bra}[1]{\ensuremath{\left\langle \right. #1 \left. \right|}}
\newcommand{\ket}[1]{\ensuremath{\left. \right| #1 \left. \right\rangle}}
\newcommand{\Av}[1]{\ensuremath{\langle #1 \rangle}}

\DeclareMathOperator{\Tr}{Tr}

\title{Vibrational Entanglement through the Lens of Quantum Information Measures}

\author{Nina Glaser}
\author{Alberto Baiardi}
\author{Annina Z. Lieberherr}
\author{Markus Reiher}
\email{mreiher@ethz.ch}
\affiliation{ETH Z\"{u}rich, Department of Chemistry and Applied Biosciences, Vladimir--Prelog-Weg 2, 8093 Z\"{u}rich, Switzerland}

\begin{document}

\begin{tocentry}
\includegraphics[width=1.\textwidth]{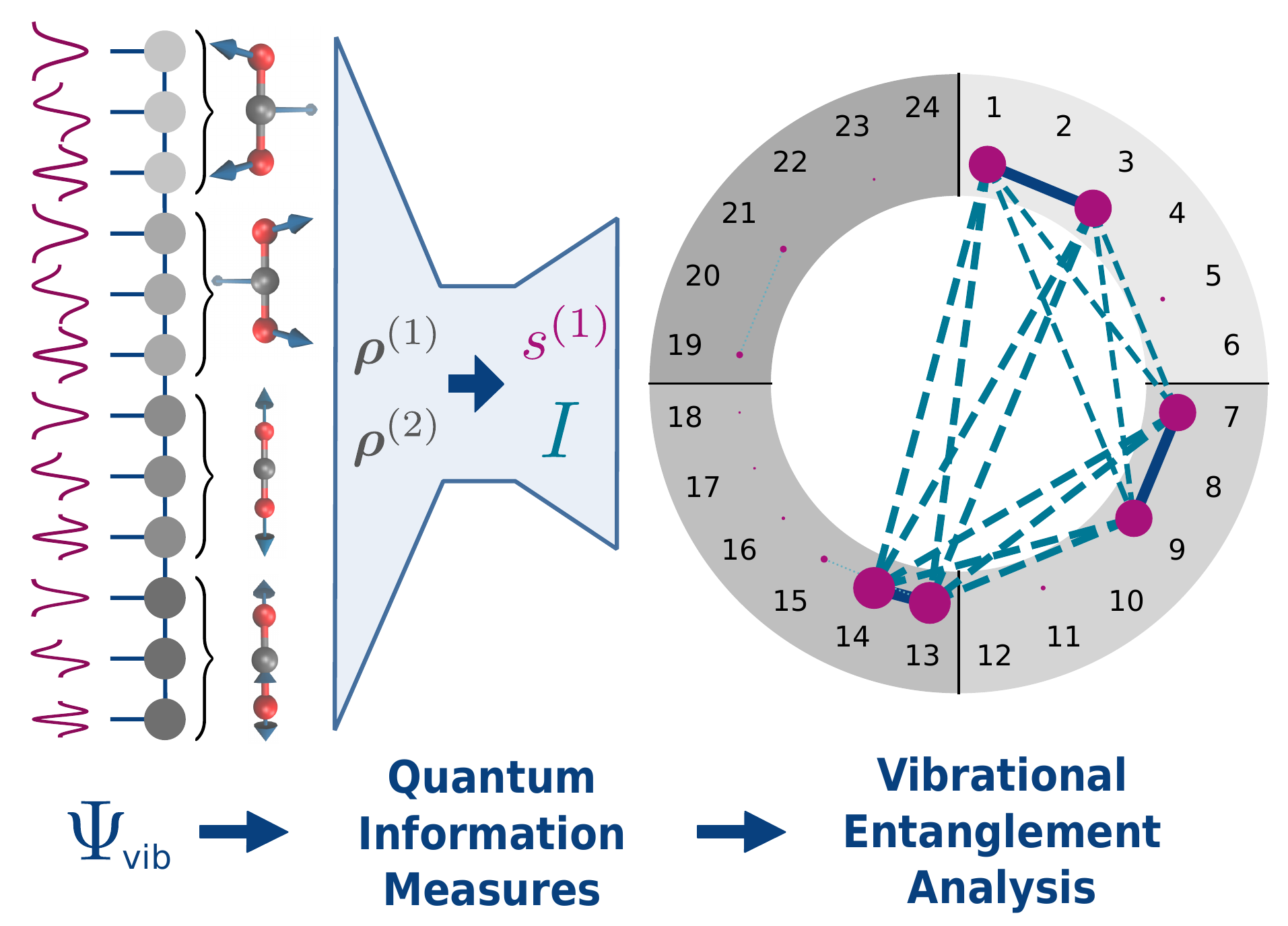}
\end{tocentry}

\begin{abstract} 
We introduce a quantum information analysis of vibrational wave functions to understand complex vibrational spectra of molecules with strong anharmonic couplings and vibrational resonances.
For this purpose, we define one- and two-modal entropies to guide the identification of strongly coupled vibrational modes and to characterize correlations within modal basis sets.
We evaluate these descriptors for multi-configurational vibrational wave functions which we calculate with the $n$-mode vibrational density matrix renormalization group algorithm.
Based on the quantum information measures, we present a vibrational entanglement analysis of the vibrational ground and excited states of CO$_2$, which display strong anharmonic effects due to the symmetry-induced and accidental (near-) degeneracies.
We investigate the entanglement signature of the Fermi resonance and discuss the maximally entangled state arising from the two degenerate bending modes.
\end{abstract}

The calculation of anharmonic vibrational spectra remains a major challenge in computational molecular spectroscopy.\cite{puzzarini19_compSpectReview,bowman22_book}
For molecules with strong anharmonic couplings, a description of the nuclear motion beyond the harmonic approximation or a mean-field treatment is essential for the accurate prediction and interpretation of their spectra.
Multi-configurational vibrational methods are required to accurately treat correlated nuclear degrees of freedom described by complex potential energy surfaces (PESs).
However, the resulting anharmonic wave functions can no longer be as straightforwardly interpreted as within the harmonic approximation.
For this reason, an assignment of the vibrational eigenstates is commonly performed, either based on harmonic reference states or, less frequently, on mean-field wave functions obtained from vibrational self-consistent field (VSCF) calculations.
However, in the case of strong anharmonicity, such an assignment of states fails to adequately characterize highly correlated vibrational states such as resonant ones.
These multi-configurational vibrational wave functions can feature a high degree of entanglement between different vibrational modes and among different basis states of a given mode.
Therefore, we propose vibrational quantum information measures for disentangling the entanglement structure of anharmonic wave functions in this letter.
We generalize the quantum information-based quantities introduced to electronic structure theory in Refs.~\citenum{Legeza2003_Entanglement,legeza04_qinf-dmrg,Rissler2006_QuantumInformationOrbitals} to vibrational wave functions.
These information entropy measures can be applied for a diagnostic analysis of anharmonic molecular vibrations, as they provide both qualitative and quantitative insights to assess vibrational entanglement.\\

To reduce artifacts in the vibrational entanglement analysis, such as from an \textit{a priori} restriction of the multi-configurational wave function to a predefined excitation rank or due to the referencing to some zeroth-order wave function, we calculate the anharmonic vibrational wave functions with the vibrational density matrix renormalization group algorithm (vDMRG),\cite{Baiardi2017_VDMRG,glaser22} because
vDMRG does not impose a fixed pre-defined truncation of the chosen Hilbert space 
and allows for a systematic convergence towards the full vibrational configuration interaction (VCI) limit.
As such, it is an ideal method for obtaining accurate anharmonic wave functions with a balanced description of correlation effects while enabling the treatment of dozens of coupled vibrational modes by virtue of its favorable scaling with system size.
In the following, we define and calculate vibrational entanglement descriptors based on the flexible $n$-mode formulation of the vDMRG algorithm.\cite{glaser23_nmode-vDMRG}
Notwithstanding that, we emphasize that these descriptors can be applied to any kind of correlated second-quantized vibrational wave function, regardless of the method chosen for solving the vibrational structure problem.
Our vibrational entanglement analysis can be performed in general modal basis sets, irrespective of the PES parametrization scheme or the choice of vibrational coordinates.\\

$n$-mode vDMRG is a full-VCI-type method, which can be applied in conjunction with general single-particle basis sets while offering full flexibility with respect to both the functional form of the PES and its expansion coordinates.
In a vDMRG calculation, the multi-configurational many-body wave function is encoded as a matrix product state (MPS) as
\begin{equation}
  \begin{aligned}
   \ket{\Psi} 
   &= \sumover{\sigma_1, \ldots, \sigma_L}{}C_{\sigma_1, \ldots, \sigma_L} \ket{\sigma_1 \cdots \sigma_L} \\
   &= \sumover{\sigma_1, \ldots, \sigma_L}{} \; \sum\limits_{a_1 ... a_{L-1 }}^{m}
    M_{1 a_1}^{\sigma_1} M_ {a_1 a_2}^{\sigma_2} \cdots M_ {a_{L-1} 1}^{\sigma_L}  \ket{\sigma_1 \cdots \sigma_L} \\
   &= \sumover{\sigma_1, \ldots, \sigma_L}{} \textbf{M}^{\sigma_1} \cdots \textbf{M}^{\sigma_L}  \ket{\sigma_1 \cdots \sigma_L} \, ,
 \end{aligned}
 \label{eq:mps}
\end{equation}
where $L$ is the overall number of single-particle basis states $\sigma_l$ mapped onto the DMRG lattice, and $\ket{\sigma_1 \cdots \sigma_L}$ represents the corresponding many-body basis state.
The $M_{a_{i-1} a_{i}}^{\sigma_i}$ are rank-three tensors with the index $\sigma_i$ labeling the basis state of lattice site $i$, where for a given value of $\sigma_i$, $\textbf{M}^{\sigma_i}$ is matrix of at most size $m \times m$.
The upper limit for the matrix dimension $m$ is commonly referred to as the bond dimension, and it is the key parameter of any DMRG calculation as it controls the compression degree of the MPS wave function.
Although an exact reconstruction of the full-VCI wave function requires the bond dimension $m$ to grow exponentially with the basis size $L$ of the system,\cite{Schollwoeck2011_Review} we showed in previous work that for vDMRG calculations, converged energies can be obtained already with rather small values of $m$, i.e., $m\leq 100$,\cite{Baiardi2017_VDMRG,Baiardi2019_HighEnergy-vDMRG,glaser23_nmode-vDMRG,glaser22} hence effectively reducing the exponential scaling of the MPS with system size to a polynomial one.\\

In our original formulation of vDMRG,\cite{Baiardi2017_VDMRG,Baiardi2019_HighEnergy-vDMRG} the vibrational wave function was restricted to vibrational Hamiltonians expressed in a harmonic oscillator basis, with the PES being represented as Taylor series around a reference geometry.
While this canonical vDMRG algorithm is an efficient method to characterize weakly anharmonic vibrational systems, it is not well suited to treat strong anharmonic effects due to its inherently limited description of the vibrational degrees of freedom, both in terms of the basis functions and the PES parametrization.
These drawbacks can be overcome by employing a more general second quantization framework which is adaptable to strongly anharmonic systems by adopting a more flexible representation of the PES.
The PES $\mathcal{V}$ can be expressed as a general many-body expansion in which terms are grouped based on the number of degrees of freedom on which they depend.
If the PES is expanded in terms of Cartesian normal mode coordinates $\boldsymbol{Q} = (Q_1, \ldots Q_M)$, the so-called $n$-mode expansion\cite{Carter1997_VSCF-CO-Adsorbed,Bowman2003_Multimode-Code,Toffoli2007_Automatic-PES} is obtained, 
\begin{equation}
\mathcal{V}(Q_1,\ldots,Q_M) = \sum\limits_{i=1}^M \mathcal{V}_1^{[i]}(Q_i) + \sum\limits_{i<j}^M \mathcal{V}_2^{[ij]}(Q_i, Q_j) + \sum\limits_{i<j<k}^M \mathcal{V}_3^{[ijk]}(Q_i, Q_j, Q_k) + \ldots \, ,
\label{eq:pes_nmode}
\end{equation}
where the terms in $\mathcal{V}_n$ depend on exactly $n$ of the $M$ normal modes, while all other coordinates remain fixed at their reference values.
The one-body term $\mathcal{V}^{[i]} \left( Q_i \right)$ represents the anharmonic variation of the PES upon change of the $i$-th coordinate.
Similarly, the two-body term $\mathcal{V}^{[ij]} \left( Q_i, Q_j \right)$ contains the variation of the potential arising from simultaneously changing two coordinates $i$ and $j$.
The one-mode contributions $\mathcal{V}^{[i]}$ and $\mathcal{V}^{[j]}$ are removed from $\mathcal{V}^{[ij]}$ to ensure that the latter quantity includes only the pure coupling between the modes.
The $n$-mode expansion becomes exact if carried out up to the $M$-th order.
However, it has been shown that even for strongly anharmonic systems an accurate representation of the PES can be obtained by truncating the expansion in Eq.~(\ref{eq:pes_nmode}) already at low orders.\cite{Carter1997_VSCF-CO-Adsorbed,Kongsted2006_NMode,manzhos06_hdmr-nn-pes,Vendrell2007_ProtonatedWater-15D}
Furthermore, the PES expansion given in Eq.~(\ref{eq:pes_nmode}) enables one to encode the potential terms in second quantization with general single-mode basis sets,\cite{Christiansen2004_nMode-SecondQuantization} allowing for basis functions which are more suited to encode anharmonicity than harmonic oscillator functions, such as the anharmonic mean-field eigenfunctions of a VSCF calculation.\cite{Gerber1986_vSCF,Bowman1986_vscf,Hansen2010_vSCF}\\

\begin{figure}
    \centering
    \includegraphics[width = 0.8\textwidth]{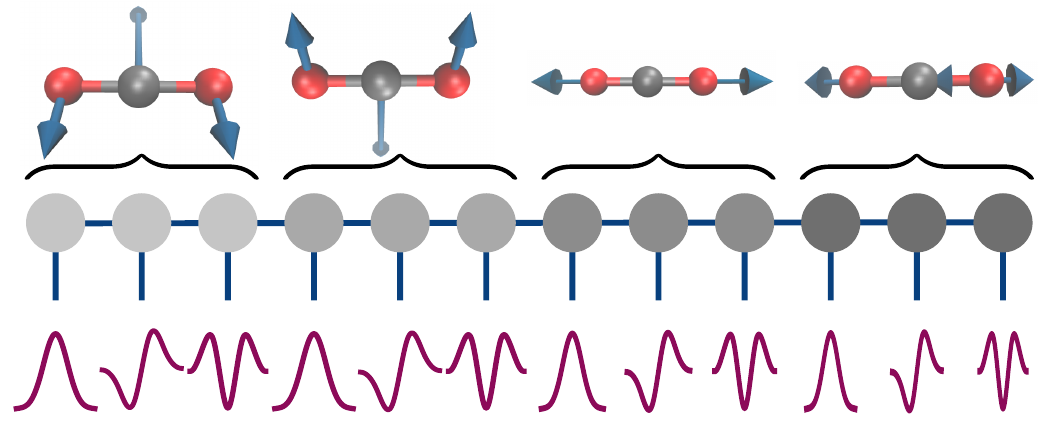}
    \caption{Illustration of the $n$-mode MPS of CO$_2$.
    In the diagrammatic tensor network notation, each filled circle denotes an MPS tensor with the adjoining lines representing the respective indices.
    For each of the four vibrational modes of CO$_2$, the three lowest-energy VSCF eigenfunctions are included as the modal basis set.}
    \label{fig:mps_co2}
\end{figure}

We denote a general single-mode basis function (commonly referred to as a modal) for mode $i$ by $\phi_i^{k_i}$, where $k_i \in \{1, \ldots, N_i\}$ with $N_i$ being the overall dimension of the basis set of a given mode.
A basis state for the full $M$-body vibrational wave function expansion can then be constructed from a product of modals as
\begin{equation}
  \psi_{k_1,\ldots, k_M} = \prod_{i=1}^M \phi_i^{k_i}(Q_i)  \, .
  \label{eq:ManyBodyBasis}
\end{equation}
The many-body basis function $\psi_{k_1,\ldots, k_M}$ can be represented as an occupation number vector (ONV) with
\begin{equation}
    \vert \mathbf{n} \rangle = | n_1^1 , \, ... \, , n^{N_1}_1, \, ... \, , n^1_i, \, ... \, , n^{N_i}_i, \, ... \, , n^1_M , \, ... \, , n^{N_M}_M \rangle \, ,
    \label{eq:onv_nmode}
\end{equation}
where $n^{k_i}_i$ denotes the occupation of the $k_i$-th modal $\phi_i^{k_i}$ associated with the $i$-th mode.
The multi-configurational wave function $\vert \Psi \rangle$ can then be efficiently encoded in $n$-mode vDMRG by mapping each modal basis function to a site on the DMRG lattice.
Hence, the $n$-mode MPS contains a tensor for each basis function included in the expansion, as is graphically illustrated in Fig.~\ref{fig:mps_co2} on the example of CO$_2$.

Based on the general many-body basis introduced in Eq.~(\ref{eq:ManyBodyBasis}), we now extend the definition of quantum information measures originally introduced in the context of electronic DMRG\cite{Legeza2003_Entanglement,legeza04_qinf-dmrg,Rissler2006_QuantumInformationOrbitals} to vibrational wave functions (\textit{cf.} also Refs. \citenum{Boguslawski2012_OrbitalEntanglement,boguslawski15_orb-ent,ding21_orb-ent} for a discussion of electron correlation and orbital entanglement).
Specifically, we define the vibrational analogs to the single- and two-orbital reduced density matrices (RDMs), the corresponding von Neumann entropies, and the orbital-pair mutual information.
Toward this end, we partition the VCI expansion in $n$-mode second quantization into a system
consisting of the $k_i$-th modal of the $i$-th mode and an environment containing all remaining basis functions.
In terms of this partition, the VCI expansion can be rewritten as
\begin{equation}
  \ket{\Psi} = \sum_{\sigma_{k_i}=0}^{1} \sum_{\bm{r}^{k_i}} C_{\sigma_{k_i} \bm{r}^{k_i}} \ket{\sigma_{k_i}} \otimes \ket{\bm{r}^{k_i}} \, ,
  \label{eq:expansion}
\end{equation}
where we collected all the modals different from the $k_i$-th one in a single ONV $\ket{\bm{r}^{k_i}}$.
The one-modal RDM can then be obtained by tracing out all possible states of $\ket{\bm{r}^{k_i}}$, \textit{i.e.}
\begin{equation}
  \bm{\rho}^{(1)}_{k_i} = \Tr_{\bm{r}^{k_i}} \ket{\Psi} \bra{\Psi} \, ,
  \label{eq:RDM_Def}
\end{equation}
where $\Tr_{\bm{r}^{k_i}}$ is the trace operator over the $\ket{\bm{r}^{k_i}}$ basis.
The one-modal RDM can be understood as an operator acting on the occupation-number basis of the $\phi_i^{k_i}$ modal.
Its matrix representation in the $\sigma_{k_i} \in \{ \ket{0}, \ket{1} \}$ basis reads
\begin{equation}
 \bm{\rho}^{(1)}_{k_i} = 
 \begin{pmatrix}
  \langle 1 - \hat{n}_i^{k_i} \rangle &                0            \\
           0                       &  \langle\hat{n}_i^{k_i}\rangle \\
 \end{pmatrix}
  \, ,
 \label{eq:1RDM_Matrix}
\end{equation}
where we introduced the number operator $\hat{n}_i^{k_i} = \bcreation{k_i} \bannihilation{k_i}$ based on the creation and annihilation operators for modal $\phi_i^{k_i}$.
For a single ONV, the diagonal elements of the one-modal RDM $\bm{\rho}^{(1)}_{k_i}$ will be either 0 and 1, if the $k_i$-th modal is occupied, or 1 and 0, if it is not contributing to the many-body state.
For a general VCI wave function the diagonal elements are between 0 and 1, and the deviation from the two extremal 
values increases with the degree of correlation of modal $\phi_i^{k_i}$ with all other ones.
This modal entanglement can be quantified with the von Neumann entropy, which we can introduce with respect to a specific basis function $\phi_i^{k_i}$ as
\begin{equation}
\begin{split}
	s_{k_i}^{(1)} &= - \Tr (\bm{\rho}^{(1)}_{k_i} \ln \bm{\rho}^{(1)}_{k_i}) \\
    &= -\sum_{\alpha} \omega_{\alpha, k_i}^{(1)} \ln \omega_{\alpha, k_i}^{(1)}  \, ,
  \label{eq:s1_definition}
\end{split}
\end{equation}
where the sum runs over the eigenvalues $\omega_{\alpha, k_i}^{(1)}$ of the RDM of the $k_i$-th basis function.
Based on the one-modal RDM given in Eq. (\ref{eq:1RDM_Matrix}), the single-modal von Neumann entropy can be conveniently calculated as
\begin{equation}
    s_{k_i}^{(1)}  = - \langle 1 - \hat{n}_i^{k_i} \rangle \ln \langle 1 - \hat{n}_i^{k_i} \rangle - \langle \hat{n}_i^{k_i} \rangle \ln \langle \hat{n}_i^{k_i} \rangle \, ,
\end{equation}
which allows for straightforward evaluation by measuring the expectation value of the occupation $\hat{n}_i^{k_i}$ of modal $\phi_i^{k_i}$ in the multi-configurational many-body wave function.
The single-modal von Neumann entropy provides a measure for the extent to which the state of the system, which here corresponds to the $k_i$-th modal, is affected by the state of the environment $\bm{r}^{k_i}$ as it quantifies the interaction of a particular modal with the modal "bath" in terms of the exchange of information.
At the same time, the single-modal entropy quantifies how much modal $\phi_i^{k_i}$ contributes to the deviation of the many-body wave function from a pure product state (\textit{i.e.}, a state where only one of the possible states $\sigma_{k_i} \in \{ \ket{0}, \ket{1} \}$ of that modal is populated).
For a pure product state, we have $s_{k_i}^{(1)}  = 0$, whereas for a maximally entangled modal we obtain $s_{k_i}^{(1)}  = \ln 2$ for $\langle \hat{n}_i^{k_i} \rangle=\frac{1}{2}$.
We note here that in the quantum computing community the von Neumann entropy is often defined as $S= - \Tr (\bm{\rho}\log_2 \bm{\rho})$,\cite{Nielsen_Chuang_2010} hence based on a base-2 logarithm, with a maximum value of $1$ for a maximally mixed single-qubit state.
However, the two definitions merely differ by a constant multiplicative factor and in terms of the resulting units (the entropy $S$ with the base-2 logarithm is commonly measured in "bits" or "shannons", whereas the natural logarithm is given in "nats"\cite{ISO_units-infScience-08}).\\

To quantify the entanglement between two specific modals, we begin by introducing the two-modal reduced density matrix.
We now consider a pair of modals $\lbrace \phi_i^{k_i}, \phi_j^{l_j} \rbrace$ as the system, where the modals can be basis functions of the same mode $(i = j)$ or belong to different modes $(i \neq j)$.
The two-modal RDM $\bm{\rho}^{(2)}_{k_i, l_j}$ is then defined as
\begin{equation}
  \bm{\rho}^{(2)}_{k_i, l_j} = \mathrm{Tr}_{\bm{r}^{k_i, l_j}} \ket{\Psi} \bra{\Psi} \, , 
  \label{eq:2RDM_Def}
\end{equation}
where we have collected all residual modals in a single ONV $\ket{\bm{r}^{k_i, l_j}}$.
The representation of $\bm{\rho}^{(2)}_{k_i, l_j}$ in the standard ONV basis $\{ \ket{00}, \ket{10}, \ket{01}, \ket{11} \}$ reads
\begin{equation}
 \bm{\rho}^{(2)}_{k_i, l_j} = 
    \begin{pmatrix}
      \langle (1 - \hat{n}_i^{k_i})(1 - \hat{n}_j^{l_j})\rangle & 0 & 0 & 0 \\
      0 & \Av{ (1 - \hat{n}_i^{k_i}) \hat{n}_j^{l_j} } & \Av{ \bannihilation{k_i} \bcreation{l_j} } & 0 \\
      0 & \Av{ \bcreation{k_i} \bannihilation{l_j} } & \Av{ \hat{n}_i^{k_i} (1 - \hat{n}_j^{l_j}) } & 0 \\
      0 & 0 & 0 & \langle \hat{n}_i^{k_i} \hat{n}_j^{l_j} \rangle
    \end{pmatrix}
    \, ,
  \label{eq:2RDM_Matrix}
\end{equation}
where the off-diagonal terms will vanish if the two modals belong to different modes as particle number symmetry is conserved.
We can now define the two-modal entropy $s^{(2)}_{k_i, l_j}$ as
\begin{equation}
  s^{(2)}_{k_i, l_j} = - \sum_{\alpha = 1}^{4} \omega^{(2)}_{\alpha,k_i l_j} \ln \omega^{(2)}_{\alpha, k_i l_j} \, ,
  \label{eq:s2_definition}
\end{equation}
where the sum runs over all eigenvalues $\omega^{(2)}_{\alpha,k_i l_j}$ of the two-modal RDM.
The two-modal entropy provides a measure for how much the combined state of two modals is affected by the environment and quantifies the information that the modal-pair possesses about the remaining modal bath and \textit{vice versa}.
If the state of the modal-pair is not correlated to the state of the residual modals at all, the two-modal entropy is zero, whereas its maximum value is bound by $s^{(2)}_{k_i, l_j}\leq s_{k_i}^{(1)} + s_{l_j}^{(1)} \leq \ln 4$.\\

To obtain the pure two-body correlation between modals $\phi_i^{k_i}$ and $\phi_j^{l_j}$, we calculate their mutual modal information which we define as
\begin{equation}
    I_{k_i, l_j} = \frac{1}{2} \left( s_{k_i}^{(1)} + s_{l_j}^{(1)} - s^{(2)}_{k_i, l_j} \right) (1 - \delta_{k_i, l_j}) \, .
      \label{eq:MutualInformation}
\end{equation}
Since the contribution of the two-modal entropy is subtracted from the single-modal entropies, the mutual modal information measures how strongly two modals are mutually correlated as all correlation with the remaining modal bath is removed.
$I_{k_i, l_j}$ evaluates to zero for any uncorrelated pair of modals, whereas its maximum value amounts to $I_{k_i, l_j}= \ln 2$ for a maximally entangled pair of modals.
The mutual information quantifies the total information one system (here modal $\phi_i^{k_i}$) has about another (modal $\phi_j^{l_j}$) and \textit{vice versa}, including all types of correlation, both classical and quantum.
Note that alternative definitions of mutual information are sometimes employed in electronic structure theory and quantum information science, which effectively, however, only differ by a constant prefactor.
Here we follow the convention that the quantum information measure will result in larger positive values if the system of interest is in a more strongly entangled state, whereas it will be lower bound by zero if the system does not possess any correlation at all.\\

The total quantum information encoded in the many-body wave function can be quantified as
\begin{equation}
    I_{\text{tot}} = \sum_{{k_i}} s_{k_i}^{(1)} \, ,
\end{equation}
where the sum runs over all single-mode basis functions of all modes.
If $I_{\text{tot}}=0$, the vibrational wave function is in a pure product state, whereas its multi-configurational character increases with larger values of $I_{\text{tot}}$.

We emphasize that while our derivation of the modal information entropies follows similar steps as for the definition of the electronic one- and two-orbital entropies,\cite{Legeza2003_Entanglement} we here account for distinguishable bosons \textit{in lieu} of indistinguishable fermions.
As a consequence, the form of the RDMs given in Eqs.~(\ref{eq:1RDM_Matrix}) and (\ref{eq:2RDM_Matrix}) is inherently different from those in the electronic case.
Furthermore, the vibrational entropies $s_{k_i}^{(1)}$ and $s^{(2)}_{k_i, l_j}$ depend on both the mode and the basis function.
We can, therefore, distinguish between inter- and intra-mode vibrational entanglement, which allows for more differentiated look at the interaction patterns encoded in complex vibrational wave functions.\\

To demonstrate our vibrational modal-entanglement analysis, we study CO$_2$ as a classical example exhibiting strong anharmonic effects due to its Fermi resonance.\cite{Fermi1931_co2-resonanz}
CO$_2$ possesses both symmetry-induced and accidental (near) degeneracies, as the two bending modes $\nu_2$ and $\bar{\nu}_2$ are degenerate, while their overtone features almost the same harmonic energy as the fundamental transition of the symmetric stretching mode $\nu_1$ so that these modes are strongly coupled.\cite{Hirata2007_CO2-FermiResonances,Hirata2008_CO2-FermiResonance-VPT}
We calculated the $n$-mode vibrational Hamiltonian of CO$_2$ in second quantization with \textsc{Colibri}\cite{colibri} by applying the computational methodology introduced in Ref.~\citenum{glaser23_nmode-vDMRG}.
For this, we obtained an accurate anharmonic PES with up to 3-body coupling terms based on explicitly correlated CCSD(T)-F12/RI electronic structure single points\cite{adler07_ccsdpart-f12} obtained in the cc-pVTZ-F12 basis\cite{peterson08_cc-pvnz-f12} with the \textsc{Orca} program\cite{neese22_orca-5} through the interface available within \textsc{Colibri}.
As a primitive basis set, we chose an 11-point discrete variable representation (DVR) basis\cite{Colbert1992_DVR} along each normal coordinate by equidistantly distributing the DVR points around the equilibrium geometry up to the 5-th harmonic inversion point of each mode.
We then transformed the second-quantized vibrational Hamiltonian into a modal basis, where we selected the six lowest-energy states of each mode as the single-particle basis set.
Several choices for the modal basis function type are available in our computational framework, namely harmonic oscillator eigenfunctions, (partially) random modal guesses, modals obtained by diagonalizing the anharmonic one-body potential, and modals calculated with our VSCF algorithm corresponding to the mean-field solution of the anharmonic Hamiltonian.
For illustrative purposes, we compare two modal basis sets in this letter, namely optimized VSCF modals and unoptimized, noisy modals constructed from the diagonalization of the anharmonic one-body potential with a 20\% uniform noise admixture.
To obtain the correlated vibrational wave function in a given modal basis, we optimized the vibrational MPS with the $n$-mode vDMRG algorithm implemented in the \textsc{QCMaquis} program.\cite{Baiardi2017_VDMRG,glaser23_nmode-vDMRG,qcmaquis_vers313}
To ensure that the multi-configurational wave function was well converged to sub-cm$^{-1}$ accuracy for the vibrational energy, the bond dimension of the MPS was dynamically adapted according to a truncation threshold of $\lambda_{\text{cut}} \leq 10^{-12}$, and the MPS was optimized until it fulfilled the chosen energy convergence threshold of $\Delta E \leq 10^{-10}$ cm$^{-1}$ from one DMRG sweep to the next.
The excited states were calculated through a constrained optimization of the MPS tensors orthogonal to all lower-lying states.
Once the MPS wave function optimization was converged, we measured the modal entropies by applying the corresponding operators in matrix form to the MPS.
The energies of the fundamental vibrations of CO$_2$ and the $\nu_2$ overtones as calculated in a VSCF modal basis are given in Table~\ref{tab:co2_freq}.

\begin{table}[htbp!]
    \centering
    \caption{
          Vibrational wavenumbers of the fundamental frequencies and the overtones of the bend vibrations of CO$_2$ in $\text{cm}^{-1}$.
          Bold font highlights the wavenumbers 
          of the Fermi doublet. $l$ denotes the vibrational angular momentum of the $\nu_2^2$ states.
          For comparison, the theoretical best estimates (TBE) calculated by Ref.~\citenum{Hirata2007_CO2-FermiResonances} and the experimental values from Ref.~\citenum{chedin79_co2_vibSpec} are listed.
          The states $\nu_2\bar{\nu}_2$ and $\nu_2^2$ ($l=2$) have not been distinguished by Hirata and coworkers, whereas we calculated and characterized both states.
          }
	\label{tab:co2_freq}
	\begin{tabular}{l | r r r r | r r }
		\hline \hline
		State & Harmonic & VSCF & VCISDT & vDMRG & TBE & Exp. \\
		\hline
    	$\nu_1$  &  1353.3 & 1341.2 & 1285.1 & \textbf{1285.1} & 1288.9 & \textbf{1285.4}\\
    	$\nu_2$    &  673.3 & 668.6 & 667.6 & 667.6 & 669.1 & 667.4\\
    	$\nu_3$   &  2394.2 & 2352.3 & 2347.0 & 2347.0 & 2349.2 & 2349.2\\
        $\nu_2^2$ ($l=0$)  & 1346.6 & 1341.3 & 1388.7 & \textbf{1388.7} & 1389.3 & \textbf{1388.2}\\
        $\nu_2\bar{\nu}_2$  & 1346.6 & 1338.4 & 1336.6 & 1336.5 & \multirow{2}{*}{1339.6} & \multirow{2}{*}{1335.1}\\
    	$\nu_2^2$ ($l=2$) & 1346.6 & 1341.3 & 1338.9 & 1336.6 &  & \\
    	\hline \hline
	\end{tabular}
\end{table}

\begin{figure}
    \centering
    \includegraphics[width = \textwidth]{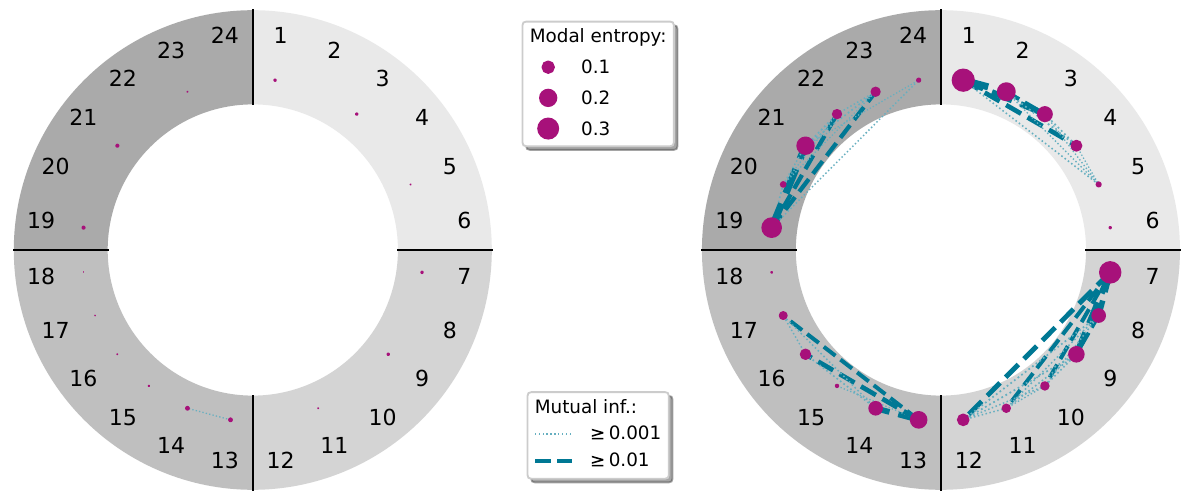}
    \caption{Modal entanglement diagrams showing the single-modal von Neumann entropies (purple circles) and the mutual modal information (connecting lines) for the vibrational ground state of CO$_2$ in an optimized VSCF modal basis (left) and an unoptimized, noisy modal basis (right).
    Each quarter of the circle is assigned to a vibrational mode, with the modes ordered with increasing energy as $\nu_2$, $\bar{\nu}_2$, $\nu_1$ and $\nu_3$ from the lighter to the darker shaded segments, respectively.
    In each segment the modal energy increases clockwise.
    }
    \label{fig:ent_zpve}
\end{figure}

For the vibrational entanglement analysis, we visualize the single-modal entropy and the mutual modal information in entanglement diagrams, similar to what has been done first for electronic structure problems.\cite{Murg10_eleStruct-TTNS,barcza11_qinf-molstruct}
For this purpose, we arrange the modals in a circular fashion, with the modals belonging to the same mode being grouped together.
The single-modal entropy and the mutual modal information are then represented as points and connecting lines, respectively, with their sizes and line strengths representing their numerical values.
Entanglement diagrams of the zero-point vibrational ground state calculated in two different modal basis sets are shown in Fig.~\ref{fig:ent_zpve}.
In these diagrams, the modals are arranged in correspondence to the MPS lattice ordering in Fig.~\ref{fig:mps_co2}, except that we now include 6 basis functions per mode.

The vibrational entanglement analysis of the ground state wave function in the optimized VSCF modal basis contains very few correlation features.
VSCF modals can encode the single-mode anharmonicity of the PES entirely, such that the resulting product state would be exact in the absence of mode-coupling terms.
They also absorb the mean-field coupling between the different modes through the VSCF procedure.
As such, the vibrational ground state wave function can be well represented with an almost pure product state, which can be seen from the very small single-modal entropies and the absence of any strong correlation as indicated by the lack of modal pairs with a large mutual modal information.

In the unoptimized noisy modal basis, however, distinct vibrational entanglement patterns can be observed.
The modals possess a significant entropy, with higher values measured for the lower energy modals than for the highly excited ones (as expected since low-energy basis functions usually contribute more to the ground state wave function than high-energy ones).
Strong correlation patterns occur within the modal basis sets of all four modes.
While the mutual modal information of modal pairs belonging to the same vibrational mode is considerably large, the inter-mode correlation is almost negligible.
This is to be expected, as the noise in the modal basis increases intra-mode entanglement, while it does not change the strength of the anharmonic couplings between the different modes.
Since both states are obtained for the same vibrational Hamiltonian, just expressed in different modal bases, the coupling between modes remains the same.
Hence, while noisy modal bases result in higher (artificial) intra-mode correlation, they do not introduce additional inter-mode correlation.
The inter-mode couplings, however, depend crucially on the expansion coordinates of the PES, meaning that a poor choice of coordinates can result in an artificially strongly coupled anharmonic Hamiltonian.
For rather rigid molecules with a well-defined equilibrium structure, such as CO$_2$, Cartesian normal modes are a natural choice for the PES expansion coordinates, but for more floppy molecules or weakly bound molecular clusters, other choices such as internal coordinates\cite{wilson55_molVib-book} or local modes\cite{Jacob2009_LocalModes} might be more suitable.
The vibrational entanglement analysis can be exploited to monitor the extrinsic correlation that arises from the use of different vibrational coordinates and modal basis sets through the inter- and intra-mode correlation patterns.
The ideal coordinate system minimizes inter-mode coupling strengths, while the optimal modal basis set minimizes the intra-mode correlation.
The total correlation can be conveniently quantified through the total quantum information contained in the system, for instance by measuring $I_{\text{tot}}$ for the vibrational ground state wave function.

\begin{figure}
    \centering
    \includegraphics[width = \textwidth]{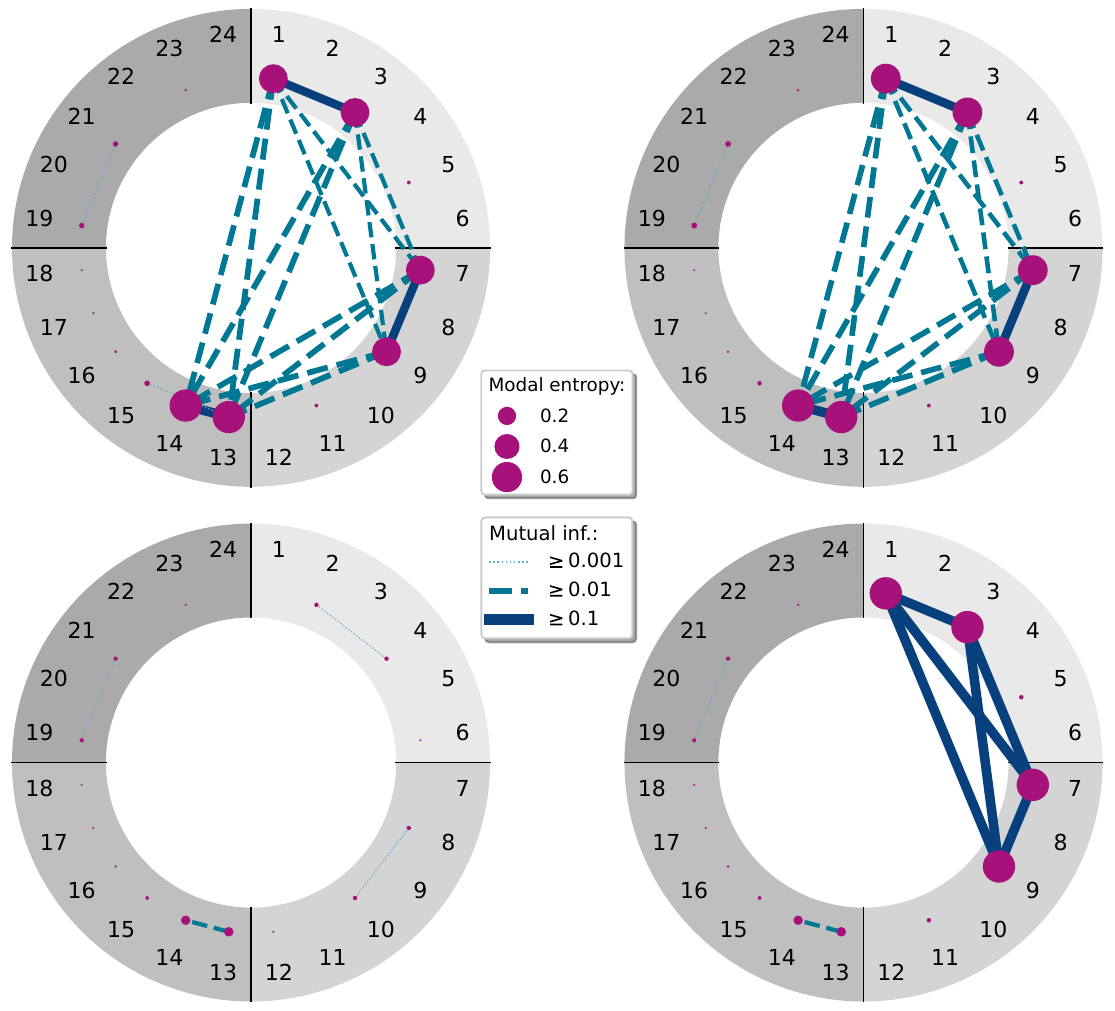}
    \caption{Modal entanglement diagrams of the vibrational states around the Fermi resonance of CO$_2$ in a VSCF modal basis.
    In the upper half, the Fermi doublet with the states $\nu_1$ (left) and $\nu_2^2$ with $l=0$ (right) is displayed, where  $l$ denotes the vibrational angular momentum quantum number of the $\nu_2^2$ overtone states.
    In the lower half, two additional near-degenerate states are shown, namely the state $\nu_2 \bar{\nu}_2$ (left) and the state $\nu_2^2$ with $l=2$ (right).}
    \label{fig:ent_res}
\end{figure}

While the vibrational ground state in an optimized VSCF modal basis is nearly uncorrelated, the entanglement analysis of the excited states corresponding to the Fermi resonance shows striking entanglement patterns, as can be seen in the upper part of Fig.~\ref{fig:ent_res}, with the Fermi doublet displaying very similar entanglement signatures.
In these resonant states, six modals have a significant entropy, namely the ground state modals of the two bending vibrations $\nu_2$ and $\bar{\nu}_2$, and the corresponding overtones, as well as the ground state modal and the fundamental excitation of the symmetric stretching mode $\nu_1$.
These modals are strongly correlated, as indicated by their considerable pairwise mutual modal information, with the largest entanglement occurring between the modals belonging to the same vibrational mode.
The total quantum information contained in these states is substantial, even though they are represented in an anharmonic, mean-field optimal modal basis.
Hence, these vibrational wave functions are strongly multi-configurational, with the large single-modal entropies distinctively characterizing the multi-configurational state of the Fermi doublet.
By analyzing the CI coefficients of the wave functions with the stochastic reconstruction of the complete active space (SRCAS) algorithm\cite{boguslawski11_srcas,glaser23_nmode-vDMRG} that reconstructs these coefficients from the vibrational MPS, we obtained $\Psi^{\text{vDMRG}}_{\nu_1} = 0.73 \Psi^{\text{VSCF}}_{\nu_1} - 0.48(\Psi^{\text{VSCF}}_{\nu_2^2} + \Psi^{\text{VSCF}}_{\bar{\nu}_2^2} )$ and $\Psi^{\text{vDMRG}}_{\nu_2^2,l=0} = 0.68 \Psi^{\text{VSCF}}_{\nu_1} + 0.52(\Psi^{\text{VSCF}}_{\nu_2^2} + \Psi^{\text{VSCF}}_{\bar{\nu}_2^2} )$.\\

In addition to the Fermi doublet, we show two further nearly degenerate excited states in Fig.~\ref{fig:ent_res}.
While these states are within the energy interval of the Fermi resonance, their wave functions transform according to different irreducible representations.
These differences in symmetry can be easily observed in the vibrational entanglement analysis through their distinct entanglement signatures.
Whereas for the state $\nu_2\bar{\nu}_2$ there is very little correlation, the state $\nu_2^2$ with $l=2$ exhibits strong entanglement between the ground state modals and the second overtones of the two degenerate bending modes.
In fact, the symmetry-induced degeneracy of these two vibrational modes results in a nearly maximally entangled state, where all four involved modals have a von Neumann entropy of $s_{k_i}^{(1)} \approx \ln 2$.
The entanglement is evenly distributed, as the mutual modal information between all strongly correlated modal pairs is almost equal, with $I_{k_i, l_j} \approx \frac{\ln 2}{4}$.
If one only considers the two involved modes $\nu_2$ and $\bar{\nu}_2$, and regards the two significant modals of each mode as the two possible states of this mode, one can draw a close analogy of this vibrational state to a Bell state.
The state of the vibrational resonance can be reduced to $\vert \Psi \rangle = \frac{1}{\sqrt{2}} \left( \vert 2 \rangle_{\nu_2} \otimes  \vert 0 \rangle_{\bar{\nu}_2} - \vert 0 \rangle_{\nu_2} \otimes  \vert 2 \rangle_{\bar{\nu}_2} \right) $ if one drops all modals except the ground and overtone ones of the two bending modes.
This corresponds exactly to the Bell state $\vert \Psi^- \rangle$, which is a maximally entangled two-qubit state.
Due to the principle of monogamy of entanglement,\cite{Osborne06_monogamyOfEnt} 
the involved modals can thus not be significantly entangled with any other modals in this state.
In fact, the mutual information of bending modes $\nu_2$ and $\bar{\nu}_2$ amounts to the maximum value $I_{\nu_2, \bar{\nu}_2}=\ln 2$.\\

In this work, we introduced modal entropy descriptors that provide both qualitative and quantitative insights into correlated wave functions to assess vibrational entanglement and correlation effects.
Degeneracies, whether symmetry-induced or accidental, result in distinct entanglement signatures, which enable an intuitive visual assessment of the wave function characteristics.
The modal entropy descriptors allow for a closer look at the multi-configurational character of a given vibrational state, and provide additional insights into the interactions encoded in a specific vibrational Hamiltonian.
Although we performed the vibrational entanglement analysis on MPS wave functions calculated with the vDMRG algorithm, the modal entropy descriptors introduced in this letter are completely general and can be easily transferred to other types of second-quantized multi-configurational vibrational methods such as those based on vibrational configuration interaction or vibrational coupled cluster.\cite{christiansen07_vibWFmethods}
As a consequence, the vibrational entanglement analysis can be applied to compare the results of different vibrational structure methods, PES parameterizations, vibrational coordinates, and modal basis sets.
The quantum information descriptors can, for instance, be utilized to monitor the introduction of artificial inter- and intra-mode correlations caused by a poor choice of the expansion coordinates and the modal basis, respectively.
The modal entropy descriptors can also be exploited to lower the computational costs of multi-configurational anharmonic calculations or to further enhance their accuracy, \textit{e.g.} by utilizing them
1) to determine the subset of strongly correlated vibrational modes that require treatment beyond the harmonic approximation or mean-field approach,
2) to select which modal basis functions to include in a multi-configurational calculation (for instance, in the spirit of the \textsc{AutoCAS} algorithm\cite{Stein2016_AutomatedSelection} for the automated selection of the active space orbitals in multi-configurational electronic structure calculations),
3) to directly optimize the modal basis set with approaches inspired by orbital optimization schemes such as the \textsc{QICAS} method\cite{ding23_qicas},
4) to optimize the ordering of both the modes and the modals on the DMRG lattice as is commonly done in electronic structure DMRG calculations,\cite{Legeza2003_Entanglement,Moritz2005_OptimalOrdering-DMRG,barcza11_qinf-molstruct}
and 5) to establish a hierarchy of strongly correlated modes to guide the construction of tree tensor network states\cite{larsson19_vibTTNS}.
We also note that, while we have applied the entanglement analysis to vibrational eigenstates in the present work, the modal entropy descriptors could also be straightforwardly exploited in time-dependent nuclear quantum dynamics simulations, for instance, to gain further insights into vibrational energy redistribution phenomena or to optimize the mode combination and layering scheme of the commonly employed multi-layer multi-configurational time-dependent Hartree \textit{ansatz}.\cite{vendrell11_mlmctdh,mendive-tapia23_modeComb}

\begin{acknowledgement}

The authors gratefully acknowledge the financial support from the Swiss National Science Foundation through grant No. 200021\_219616.

\end{acknowledgement}

\begin{mcitethebibliography}{49}
\providecommand*\natexlab[1]{#1}
\providecommand*\mciteSetBstSublistMode[1]{}
\providecommand*\mciteSetBstMaxWidthForm[2]{}
\providecommand*\mciteBstWouldAddEndPuncttrue
  {\def\EndOfBibitem{\unskip.}}
\providecommand*\mciteBstWouldAddEndPunctfalse
  {\let\EndOfBibitem\relax}
\providecommand*\mciteSetBstMidEndSepPunct[3]{}
\providecommand*\mciteSetBstSublistLabelBeginEnd[3]{}
\providecommand*\EndOfBibitem{}
\mciteSetBstSublistMode{f}
\mciteSetBstMaxWidthForm{subitem}{(\alph{mcitesubitemcount})}
\mciteSetBstSublistLabelBeginEnd
  {\mcitemaxwidthsubitemform\space}
  {\relax}
  {\relax}

\bibitem[Puzzarini \latin{et~al.}(2019)Puzzarini, Bloino, Tasinato, and
  Barone]{puzzarini19_compSpectReview}
Puzzarini,~C.; Bloino,~J.; Tasinato,~N.; Barone,~V. {Accuracy and
  Interpretability: The Devil and the Holy Grail. New Routes across Old
  Boundaries in Computational Spectroscopy}. \emph{Chem. Rev.} \textbf{2019},
  \emph{119}, 8131--8191\relax
\mciteBstWouldAddEndPuncttrue
\mciteSetBstMidEndSepPunct{\mcitedefaultmidpunct}
{\mcitedefaultendpunct}{\mcitedefaultseppunct}\relax
\EndOfBibitem
\bibitem[Bowman(2022)]{bowman22_book}
Bowman,~J.~M., Ed. \emph{{Vibrational Dynamics of Molecules}}; WORLD
  SCIENTIFIC, 2022\relax
\mciteBstWouldAddEndPuncttrue
\mciteSetBstMidEndSepPunct{\mcitedefaultmidpunct}
{\mcitedefaultendpunct}{\mcitedefaultseppunct}\relax
\EndOfBibitem
\bibitem[Legeza and S\'oloyom(2003)Legeza, and
  S\'oloyom]{Legeza2003_Entanglement}
Legeza,~{\"O}.; S\'oloyom,~J. {Optimizing the density-matrix renormalization
  group method using quantum information entropy}. \emph{Phys. Rev. B}
  \textbf{2003}, \emph{68}, 195116\relax
\mciteBstWouldAddEndPuncttrue
\mciteSetBstMidEndSepPunct{\mcitedefaultmidpunct}
{\mcitedefaultendpunct}{\mcitedefaultseppunct}\relax
\EndOfBibitem
\bibitem[Legeza and S\'olyom(2004)Legeza, and S\'olyom]{legeza04_qinf-dmrg}
Legeza,~O.; S\'olyom,~J. {Quantum data compression, quantum information
  generation, and the density-matrix renormalization-group method}. \emph{Phys.
  Rev. B} \textbf{2004}, \emph{70}, 205118\relax
\mciteBstWouldAddEndPuncttrue
\mciteSetBstMidEndSepPunct{\mcitedefaultmidpunct}
{\mcitedefaultendpunct}{\mcitedefaultseppunct}\relax
\EndOfBibitem
\bibitem[Rissler \latin{et~al.}(2006)Rissler, Noack, and
  White]{Rissler2006_QuantumInformationOrbitals}
Rissler,~J.; Noack,~R.~M.; White,~S.~R. {Measuring orbital interaction using
  quantum information theory}. \emph{Chem. Phys.} \textbf{2006}, \emph{323},
  519--531\relax
\mciteBstWouldAddEndPuncttrue
\mciteSetBstMidEndSepPunct{\mcitedefaultmidpunct}
{\mcitedefaultendpunct}{\mcitedefaultseppunct}\relax
\EndOfBibitem
\bibitem[Baiardi \latin{et~al.}(2017)Baiardi, Stein, Barone, and
  Reiher]{Baiardi2017_VDMRG}
Baiardi,~A.; Stein,~C.~J.; Barone,~V.; Reiher,~M. {Vibrational Density Matrix
  Renormalization Group}. \emph{J. Chem. Theory Comput.} \textbf{2017},
  \emph{13}, 3764--3777\relax
\mciteBstWouldAddEndPuncttrue
\mciteSetBstMidEndSepPunct{\mcitedefaultmidpunct}
{\mcitedefaultendpunct}{\mcitedefaultseppunct}\relax
\EndOfBibitem
\bibitem[{Glaser} \latin{et~al.}(2022){Glaser}, {Baiardi}, and
  {Reiher}]{glaser22}
{Glaser},~N.; {Baiardi},~A.; {Reiher},~M. In \emph{Vibrational Dynamics of
  Molecules}; Bowman,~J.~M., Ed.; World Scientific, 2022; Chapter 3, pp
  80--144\relax
\mciteBstWouldAddEndPuncttrue
\mciteSetBstMidEndSepPunct{\mcitedefaultmidpunct}
{\mcitedefaultendpunct}{\mcitedefaultseppunct}\relax
\EndOfBibitem
\bibitem[Glaser \latin{et~al.}(2023)Glaser, Baiardi, and
  Reiher]{glaser23_nmode-vDMRG}
Glaser,~N.; Baiardi,~A.; Reiher,~M. {Flexible DMRG-Based Framework for
  Anharmonic Vibrational Calculations}. \emph{J. Chem. Theory Comput.}
  \textbf{2023}, \emph{19}, 9329--9343\relax
\mciteBstWouldAddEndPuncttrue
\mciteSetBstMidEndSepPunct{\mcitedefaultmidpunct}
{\mcitedefaultendpunct}{\mcitedefaultseppunct}\relax
\EndOfBibitem
\bibitem[Schollw\"ock(2011)]{Schollwoeck2011_Review}
Schollw\"ock,~U. {The density-matrix renormalization group in the age of matrix
  product states}. \emph{Ann. Phys.} \textbf{2011}, \emph{326}, 96--192\relax
\mciteBstWouldAddEndPuncttrue
\mciteSetBstMidEndSepPunct{\mcitedefaultmidpunct}
{\mcitedefaultendpunct}{\mcitedefaultseppunct}\relax
\EndOfBibitem
\bibitem[Baiardi \latin{et~al.}(2019)Baiardi, Stein, Barone, and
  Reiher]{Baiardi2019_HighEnergy-vDMRG}
Baiardi,~A.; Stein,~C.~J.; Barone,~V.; Reiher,~M. {Optimization of highly
  excited matrix product states with an application to vibrational
  spectroscopy}. \emph{J. Chem. Phys.} \textbf{2019}, \emph{150}, 094113\relax
\mciteBstWouldAddEndPuncttrue
\mciteSetBstMidEndSepPunct{\mcitedefaultmidpunct}
{\mcitedefaultendpunct}{\mcitedefaultseppunct}\relax
\EndOfBibitem
\bibitem[Carter \latin{et~al.}(1997)Carter, Culik, and
  Bowman]{Carter1997_VSCF-CO-Adsorbed}
Carter,~S.; Culik,~S.~J.; Bowman,~J.~M. {Vibrational self-consistent field
  method for many-mode systems: A new approach and application to the
  vibrations of CO adsorbed on Cu(100)}. \emph{J. Chem. Phys.} \textbf{1997},
  \emph{107}, 10458--10469\relax
\mciteBstWouldAddEndPuncttrue
\mciteSetBstMidEndSepPunct{\mcitedefaultmidpunct}
{\mcitedefaultendpunct}{\mcitedefaultseppunct}\relax
\EndOfBibitem
\bibitem[Bowman \latin{et~al.}(2003)Bowman, Carter, and
  Huang]{Bowman2003_Multimode-Code}
Bowman,~J.~M.; Carter,~S.; Huang,~X. {MULTIMODE: A code to calculate
  rovibrational energies of polyatomic molecules}. \emph{Int. Rev. Phys. Chem.}
  \textbf{2003}, \emph{22}, 533--549\relax
\mciteBstWouldAddEndPuncttrue
\mciteSetBstMidEndSepPunct{\mcitedefaultmidpunct}
{\mcitedefaultendpunct}{\mcitedefaultseppunct}\relax
\EndOfBibitem
\bibitem[Toffoli \latin{et~al.}(2007)Toffoli, Kongsted, and
  Christiansen]{Toffoli2007_Automatic-PES}
Toffoli,~D.; Kongsted,~J.; Christiansen,~O. {Automatic generation of potential
  energy and property surfaces of polyatomic molecules in normal coordinates}.
  \emph{J. Chem. Phys.} \textbf{2007}, \emph{127}, 204016\relax
\mciteBstWouldAddEndPuncttrue
\mciteSetBstMidEndSepPunct{\mcitedefaultmidpunct}
{\mcitedefaultendpunct}{\mcitedefaultseppunct}\relax
\EndOfBibitem
\bibitem[Kongsted and Christiansen(2006)Kongsted, and
  Christiansen]{Kongsted2006_NMode}
Kongsted,~J.; Christiansen,~O. {Automatic generation of force fields and
  property surfaces for use in variational vibrational calculations of
  anharmonic vibrational energies and zero-point vibrational averaged
  properties}. \emph{J. Chem. Phys.} \textbf{2006}, \emph{125}, 124108\relax
\mciteBstWouldAddEndPuncttrue
\mciteSetBstMidEndSepPunct{\mcitedefaultmidpunct}
{\mcitedefaultendpunct}{\mcitedefaultseppunct}\relax
\EndOfBibitem
\bibitem[Manzhos and Carrington(2006)Manzhos, and
  Carrington]{manzhos06_hdmr-nn-pes}
Manzhos,~S.; Carrington,~T.,~Jr. {A random-sampling high dimensional model
  representation neural network for building potential energy surfaces}.
  \emph{J. Chem. Phys.} \textbf{2006}, \emph{125}, 084109\relax
\mciteBstWouldAddEndPuncttrue
\mciteSetBstMidEndSepPunct{\mcitedefaultmidpunct}
{\mcitedefaultendpunct}{\mcitedefaultseppunct}\relax
\EndOfBibitem
\bibitem[Vendrell \latin{et~al.}(2007)Vendrell, Gatti, Lauvergnat, and
  Meyer]{Vendrell2007_ProtonatedWater-15D}
Vendrell,~O.; Gatti,~F.; Lauvergnat,~D.; Meyer,~H.-D. {Full-dimensional
  (15-dimensional) quantum-dynamical simulation of the protonated water dimer.
  I. Hamiltonian setup and analysis of the ground vibrational state}. \emph{J.
  Chem. Phys.} \textbf{2007}, \emph{127}, 184302\relax
\mciteBstWouldAddEndPuncttrue
\mciteSetBstMidEndSepPunct{\mcitedefaultmidpunct}
{\mcitedefaultendpunct}{\mcitedefaultseppunct}\relax
\EndOfBibitem
\bibitem[Christiansen(2004)]{Christiansen2004_nMode-SecondQuantization}
Christiansen,~O. A second quantization formulation of multimode dynamics.
  \emph{J. Chem. Phys.} \textbf{2004}, \emph{120}, 2140--2148\relax
\mciteBstWouldAddEndPuncttrue
\mciteSetBstMidEndSepPunct{\mcitedefaultmidpunct}
{\mcitedefaultendpunct}{\mcitedefaultseppunct}\relax
\EndOfBibitem
\bibitem[Ratner and Gerber(1986)Ratner, and Gerber]{Gerber1986_vSCF}
Ratner,~M.~A.; Gerber,~R.~B. {Excited vibrational states of polyatomic
  molecules: the semiclassical self-consistent field approach}. \emph{J. Phys.
  Chem.} \textbf{1986}, \emph{90}, 20--30\relax
\mciteBstWouldAddEndPuncttrue
\mciteSetBstMidEndSepPunct{\mcitedefaultmidpunct}
{\mcitedefaultendpunct}{\mcitedefaultseppunct}\relax
\EndOfBibitem
\bibitem[Bowman(1986)]{Bowman1986_vscf}
Bowman,~J.~M. {The self-consistent-field approach to polyatomic vibrations}.
  \emph{Acc. Chem. Res.} \textbf{1986}, \emph{19}, 202\relax
\mciteBstWouldAddEndPuncttrue
\mciteSetBstMidEndSepPunct{\mcitedefaultmidpunct}
{\mcitedefaultendpunct}{\mcitedefaultseppunct}\relax
\EndOfBibitem
\bibitem[Hansen \latin{et~al.}(2010)Hansen, Sparta, Seidler, Toffoli, and
  Christiansen]{Hansen2010_vSCF}
Hansen,~M.~B.; Sparta,~M.; Seidler,~P.; Toffoli,~D.; Christiansen,~O. {New
  formulation and implementation of vibrational self-consistent field theory}.
  \emph{J. Chem. Theory Comput.} \textbf{2010}, \emph{6}, 235--248\relax
\mciteBstWouldAddEndPuncttrue
\mciteSetBstMidEndSepPunct{\mcitedefaultmidpunct}
{\mcitedefaultendpunct}{\mcitedefaultseppunct}\relax
\EndOfBibitem
\bibitem[Boguslawski \latin{et~al.}(2012)Boguslawski, Tecmer, Legeza, and
  Reiher]{Boguslawski2012_OrbitalEntanglement}
Boguslawski,~K.; Tecmer,~P.; Legeza,~{\"{O}}.; Reiher,~M. {Entanglement
  Measures for Single- and Multireference Correlation Effects}. \emph{J. Phys.
  Chem. Lett.} \textbf{2012}, \emph{3}, 3129--3135\relax
\mciteBstWouldAddEndPuncttrue
\mciteSetBstMidEndSepPunct{\mcitedefaultmidpunct}
{\mcitedefaultendpunct}{\mcitedefaultseppunct}\relax
\EndOfBibitem
\bibitem[Boguslawski and Tecmer(2015)Boguslawski, and
  Tecmer]{boguslawski15_orb-ent}
Boguslawski,~K.; Tecmer,~P. {Orbital entanglement in quantum chemistry}.
  \emph{Int. J. Quantum Chem.} \textbf{2015}, \emph{115}, 1289--1295\relax
\mciteBstWouldAddEndPuncttrue
\mciteSetBstMidEndSepPunct{\mcitedefaultmidpunct}
{\mcitedefaultendpunct}{\mcitedefaultseppunct}\relax
\EndOfBibitem
\bibitem[Ding \latin{et~al.}(2021)Ding, Mardazad, Das, Szalay, Schollwöck,
  Zimborás, and Schilling]{ding21_orb-ent}
Ding,~L.; Mardazad,~S.; Das,~S.; Szalay,~S.; Schollwöck,~U.; Zimborás,~Z.;
  Schilling,~C. {Concept of Orbital Entanglement and Correlation in Quantum
  Chemistry}. \emph{J. Chem. Theory Comput.} \textbf{2021}, \emph{17},
  79--95\relax
\mciteBstWouldAddEndPuncttrue
\mciteSetBstMidEndSepPunct{\mcitedefaultmidpunct}
{\mcitedefaultendpunct}{\mcitedefaultseppunct}\relax
\EndOfBibitem
\bibitem[Nielsen and Chuang(2010)Nielsen, and Chuang]{Nielsen_Chuang_2010}
Nielsen,~M.~A.; Chuang,~I.~L. \emph{{Quantum Computation and Quantum
  Information}}, 10th ed.; Cambridge University Press, 2010\relax
\mciteBstWouldAddEndPuncttrue
\mciteSetBstMidEndSepPunct{\mcitedefaultmidpunct}
{\mcitedefaultendpunct}{\mcitedefaultseppunct}\relax
\EndOfBibitem
\bibitem[IEC 80000-13:2008(2008)]{ISO_units-infScience-08}
\emph{{Quantities and units Part 13: Information science and technology}};
  Standard, 2008; Vol. 2008\relax
\mciteBstWouldAddEndPuncttrue
\mciteSetBstMidEndSepPunct{\mcitedefaultmidpunct}
{\mcitedefaultendpunct}{\mcitedefaultseppunct}\relax
\EndOfBibitem
\bibitem[Fermi(1931)]{Fermi1931_co2-resonanz}
Fermi,~E. {{\"U}ber den Ramaneffekt des Kohlendioxyds}. \emph{Z. Phys.}
  \textbf{1931}, \emph{71}, 250--259\relax
\mciteBstWouldAddEndPuncttrue
\mciteSetBstMidEndSepPunct{\mcitedefaultmidpunct}
{\mcitedefaultendpunct}{\mcitedefaultseppunct}\relax
\EndOfBibitem
\bibitem[Rodriguez-Garcia \latin{et~al.}(2007)Rodriguez-Garcia, Hirata, Yagi,
  Hirao, Taketsugu, Schweigert, and Tasumi]{Hirata2007_CO2-FermiResonances}
Rodriguez-Garcia,~V.; Hirata,~S.; Yagi,~K.; Hirao,~K.; Taketsugu,~T.;
  Schweigert,~I.; Tasumi,~M. {Fermi resonance in CO2: A combined electronic
  coupled-cluster and vibrational configuration-interaction prediction}.
  \emph{J. Chem. Phys.} \textbf{2007}, \emph{126}, 124303\relax
\mciteBstWouldAddEndPuncttrue
\mciteSetBstMidEndSepPunct{\mcitedefaultmidpunct}
{\mcitedefaultendpunct}{\mcitedefaultseppunct}\relax
\EndOfBibitem
\bibitem[Yagi \latin{et~al.}(2008)Yagi, Hirata, and
  Hirao]{Hirata2008_CO2-FermiResonance-VPT}
Yagi,~K.; Hirata,~S.; Hirao,~K. {Vibrational quasi-degenerate perturbation
  theory: applications to fermi resonance in CO2, H2CO, and C6H6}. \emph{Phys.
  Chem. Chem. Phys.} \textbf{2008}, \emph{10}, 1781\relax
\mciteBstWouldAddEndPuncttrue
\mciteSetBstMidEndSepPunct{\mcitedefaultmidpunct}
{\mcitedefaultendpunct}{\mcitedefaultseppunct}\relax
\EndOfBibitem
\bibitem[Glaser \latin{et~al.}(2023)Glaser, Baiardi, Kelemen, and
  Reiher]{colibri}
Glaser,~N.; Baiardi,~A.; Kelemen,~A.~K.; Reiher,~M. qcscine/colibri: Release
  1.0.0. 2023; \url{https://doi.org/10.5281/zenodo.10276683}, Development
  version\relax
\mciteBstWouldAddEndPuncttrue
\mciteSetBstMidEndSepPunct{\mcitedefaultmidpunct}
{\mcitedefaultendpunct}{\mcitedefaultseppunct}\relax
\EndOfBibitem
\bibitem[Adler \latin{et~al.}(2007)Adler, Knizia, and
  Werner]{adler07_ccsdpart-f12}
Adler,~T.~B.; Knizia,~G.; Werner,~H.-J. {A simple and efficient CCSD(T)-F12
  approximation}. \emph{J. Chem. Phys.} \textbf{2007}, \emph{127}, 221106\relax
\mciteBstWouldAddEndPuncttrue
\mciteSetBstMidEndSepPunct{\mcitedefaultmidpunct}
{\mcitedefaultendpunct}{\mcitedefaultseppunct}\relax
\EndOfBibitem
\bibitem[Peterson \latin{et~al.}(2008)Peterson, Adler, and
  Werner]{peterson08_cc-pvnz-f12}
Peterson,~K.~A.; Adler,~T.~B.; Werner,~H.-J. {Systematically convergent basis
  sets for explicitly correlated wavefunctions: The atoms H, He, B–Ne, and
  Al–Ar}. \emph{J. Chem. Phys.} \textbf{2008}, \emph{128}, 084102\relax
\mciteBstWouldAddEndPuncttrue
\mciteSetBstMidEndSepPunct{\mcitedefaultmidpunct}
{\mcitedefaultendpunct}{\mcitedefaultseppunct}\relax
\EndOfBibitem
\bibitem[Neese(2022)]{neese22_orca-5}
Neese,~F. {Software update: The ORCA program system—Version 5.0}. \emph{WIREs
  Comput. Mol. Sci.} \textbf{2022}, \emph{12}, e1606\relax
\mciteBstWouldAddEndPuncttrue
\mciteSetBstMidEndSepPunct{\mcitedefaultmidpunct}
{\mcitedefaultendpunct}{\mcitedefaultseppunct}\relax
\EndOfBibitem
\bibitem[Colbert and Miller(1992)Colbert, and Miller]{Colbert1992_DVR}
Colbert,~D.~T.; Miller,~W.~H. {A novel discrete variable representation for
  quantum mechanical reactive scattering via the S‐matrix Kohn method}.
  \emph{J. Chem. Phys.} \textbf{1992}, \emph{96}, 1982--1991\relax
\mciteBstWouldAddEndPuncttrue
\mciteSetBstMidEndSepPunct{\mcitedefaultmidpunct}
{\mcitedefaultendpunct}{\mcitedefaultseppunct}\relax
\EndOfBibitem
\bibitem[Baiardi \latin{et~al.}(2023)Baiardi, Battaglia, Feldmann, Freitag,
  Glaser, Kelemen, Keller, Knecht, Ma, Stein, and Reiher]{qcmaquis_vers313}
Baiardi,~A.; Battaglia,~S.; Feldmann,~R.; Freitag,~L.; Glaser,~N.;
  Kelemen,~A.~K.; Keller,~S.; Knecht,~S.; Ma,~Y.; Stein,~C.~J. \latin{et~al.}
  qcscine/qcmaquis: Release 3.1.3. 2023;
  \url{https://doi.org/10.5281/zenodo.7907785}, Development version\relax
\mciteBstWouldAddEndPuncttrue
\mciteSetBstMidEndSepPunct{\mcitedefaultmidpunct}
{\mcitedefaultendpunct}{\mcitedefaultseppunct}\relax
\EndOfBibitem
\bibitem[Chedin(1979)]{chedin79_co2_vibSpec}
Chedin,~A. {The carbon dioxide molecule: Potential, spectroscopic, and
  molecular constants from its infrared spectrum}. \emph{J. Mol. Spectrosc.}
  \textbf{1979}, \emph{76}, 430--491\relax
\mciteBstWouldAddEndPuncttrue
\mciteSetBstMidEndSepPunct{\mcitedefaultmidpunct}
{\mcitedefaultendpunct}{\mcitedefaultseppunct}\relax
\EndOfBibitem
\bibitem[Murg \latin{et~al.}(2010)Murg, Verstraete, Legeza, and
  Noack]{Murg10_eleStruct-TTNS}
Murg,~V.; Verstraete,~F.; Legeza,~O.; Noack,~R.~M. {Simulating strongly
  correlated quantum systems with tree tensor networks}. \emph{Phys. Rev. B}
  \textbf{2010}, \emph{82}, 205105\relax
\mciteBstWouldAddEndPuncttrue
\mciteSetBstMidEndSepPunct{\mcitedefaultmidpunct}
{\mcitedefaultendpunct}{\mcitedefaultseppunct}\relax
\EndOfBibitem
\bibitem[Barcza \latin{et~al.}(2011)Barcza, Legeza, Marti, and
  Reiher]{barcza11_qinf-molstruct}
Barcza,~G.; Legeza,~O.; Marti,~K.~H.; Reiher,~M. {Quantum-information analysis
  of electronic states of different molecular structures}. \emph{Phys. Rev. A}
  \textbf{2011}, \emph{83}, 012508\relax
\mciteBstWouldAddEndPuncttrue
\mciteSetBstMidEndSepPunct{\mcitedefaultmidpunct}
{\mcitedefaultendpunct}{\mcitedefaultseppunct}\relax
\EndOfBibitem
\bibitem[Wilson \latin{et~al.}()Wilson, Decius, and
  Cross]{wilson55_molVib-book}
Wilson,~J.,~E.~Bright; Decius,~J.~G.; Cross,~P.~G. \emph{{Molecular Vibrations:
  The Theory of Infrared and Raman Vibrational Spectra}}; McGraw--Hill, Vol.
  102\relax
\mciteBstWouldAddEndPuncttrue
\mciteSetBstMidEndSepPunct{\mcitedefaultmidpunct}
{\mcitedefaultendpunct}{\mcitedefaultseppunct}\relax
\EndOfBibitem
\bibitem[Jacob and Reiher(2009)Jacob, and Reiher]{Jacob2009_LocalModes}
Jacob,~C.~R.; Reiher,~M. {Localizing normal modes in large molecules}. \emph{J.
  Chem. Phys.} \textbf{2009}, \emph{130}, 84106\relax
\mciteBstWouldAddEndPuncttrue
\mciteSetBstMidEndSepPunct{\mcitedefaultmidpunct}
{\mcitedefaultendpunct}{\mcitedefaultseppunct}\relax
\EndOfBibitem
\bibitem[Boguslawski \latin{et~al.}(2011)Boguslawski, Marti, and
  Reiher]{boguslawski11_srcas}
Boguslawski,~K.; Marti,~K.~H.; Reiher,~M. {Construction of CASCI-type wave
  functions for very large active spaces}. \emph{J. Chem. Phys.} \textbf{2011},
  \emph{134}, 224101\relax
\mciteBstWouldAddEndPuncttrue
\mciteSetBstMidEndSepPunct{\mcitedefaultmidpunct}
{\mcitedefaultendpunct}{\mcitedefaultseppunct}\relax
\EndOfBibitem
\bibitem[Osborne and Verstraete(2006)Osborne, and
  Verstraete]{Osborne06_monogamyOfEnt}
Osborne,~T.~J.; Verstraete,~F. {General Monogamy Inequality for Bipartite Qubit
  Entanglement}. \emph{Phys. Rev. Lett.} \textbf{2006}, \emph{96}, 220503\relax
\mciteBstWouldAddEndPuncttrue
\mciteSetBstMidEndSepPunct{\mcitedefaultmidpunct}
{\mcitedefaultendpunct}{\mcitedefaultseppunct}\relax
\EndOfBibitem
\bibitem[Christiansen(2007)]{christiansen07_vibWFmethods}
Christiansen,~O. {Vibrational structure theory: new vibrational wave function
  methods for calculation of anharmonic vibrational energies and vibrational
  contributions to molecular properties}. \emph{Phys. Chem. Chem. Phys.}
  \textbf{2007}, \emph{9}, 2942--2953\relax
\mciteBstWouldAddEndPuncttrue
\mciteSetBstMidEndSepPunct{\mcitedefaultmidpunct}
{\mcitedefaultendpunct}{\mcitedefaultseppunct}\relax
\EndOfBibitem
\bibitem[Stein and Reiher(2016)Stein, and Reiher]{Stein2016_AutomatedSelection}
Stein,~C.~J.; Reiher,~M. {Automated Selection of Active Orbital Spaces}.
  \emph{J. Chem. Theory Comput.} \textbf{2016}, \emph{12}, 1760--1771\relax
\mciteBstWouldAddEndPuncttrue
\mciteSetBstMidEndSepPunct{\mcitedefaultmidpunct}
{\mcitedefaultendpunct}{\mcitedefaultseppunct}\relax
\EndOfBibitem
\bibitem[Ding \latin{et~al.}(2023)Ding, Knecht, and Schilling]{ding23_qicas}
Ding,~L.; Knecht,~S.; Schilling,~C. {Quantum Information-Assisted Complete
  Active Space Optimization (QICAS)}. \emph{J. Phys. Chem. Lett.}
  \textbf{2023}, \emph{14}, 11022--11029\relax
\mciteBstWouldAddEndPuncttrue
\mciteSetBstMidEndSepPunct{\mcitedefaultmidpunct}
{\mcitedefaultendpunct}{\mcitedefaultseppunct}\relax
\EndOfBibitem
\bibitem[Moritz \latin{et~al.}(2005)Moritz, Hess, and
  Reiher]{Moritz2005_OptimalOrdering-DMRG}
Moritz,~G.; Hess,~B.~A.; Reiher,~M. {Convergence behavior of the density-matrix
  renormalization group algorithm for optimized orbital orderings}. \emph{J.
  Chem. Phys.} \textbf{2005}, \emph{122}, 024107\relax
\mciteBstWouldAddEndPuncttrue
\mciteSetBstMidEndSepPunct{\mcitedefaultmidpunct}
{\mcitedefaultendpunct}{\mcitedefaultseppunct}\relax
\EndOfBibitem
\bibitem[Larsson(2019)]{larsson19_vibTTNS}
Larsson,~H.~R. {Computing vibrational eigenstates with tree tensor network
  states (TTNS)}. \emph{J. Chem. Phys.} \textbf{2019}, \emph{151}, 204102\relax
\mciteBstWouldAddEndPuncttrue
\mciteSetBstMidEndSepPunct{\mcitedefaultmidpunct}
{\mcitedefaultendpunct}{\mcitedefaultseppunct}\relax
\EndOfBibitem
\bibitem[Vendrell and Meyer(2011)Vendrell, and Meyer]{vendrell11_mlmctdh}
Vendrell,~O.; Meyer,~H.-D. {Multilayer multiconfiguration time-dependent
  Hartree method: Implementation and applications to a Henon–Heiles
  Hamiltonian and to pyrazine}. \emph{J. Chem. Phys.} \textbf{2011},
  \emph{134}, 044135\relax
\mciteBstWouldAddEndPuncttrue
\mciteSetBstMidEndSepPunct{\mcitedefaultmidpunct}
{\mcitedefaultendpunct}{\mcitedefaultseppunct}\relax
\EndOfBibitem
\bibitem[Mendive-Tapia \latin{et~al.}(2023)Mendive-Tapia, Meyer, and
  Vendrell]{mendive-tapia23_modeComb}
Mendive-Tapia,~D.; Meyer,~H.-D.; Vendrell,~O. Optimal Mode Combination in the
  Multiconfiguration Time-Dependent Hartree Method through Multivariate
  Statistics: Factor Analysis and Hierarchical Clustering. \emph{J. Chem.
  Theory Comput.} \textbf{2023}, \emph{19}, 1144--1156\relax
\mciteBstWouldAddEndPuncttrue
\mciteSetBstMidEndSepPunct{\mcitedefaultmidpunct}
{\mcitedefaultendpunct}{\mcitedefaultseppunct}\relax
\EndOfBibitem
\end{mcitethebibliography}
\providecommand{\latin}[1]{#1}
\makeatletter
\providecommand{\doi}
  {\begingroup\let\do\@makeother\dospecials
  \catcode`\{=1 \catcode`\}=2 \doi@aux}
\providecommand{\doi@aux}[1]{\endgroup\texttt{#1}}
\makeatother
\providecommand*\mcitethebibliography{\thebibliography}
\csname @ifundefined\endcsname{endmcitethebibliography}
  {\let\endmcitethebibliography\endthebibliography}{}

\end{document}